\documentclass[superscriptaddress,reprint,longbibliography]{revtex4-1}
\usepackage{amsmath,amssymb,graphicx,hyperref}

\begin{document}

\title{Nonadditivity of Critical Casimir Forces}

\author{Sathyanarayana Paladugu}
\thanks{Equal contributors}
\affiliation{Soft Matter Lab, Department of Physics, Bilkent University, Ankara 06800, Turkey}

\author{Agnese Callegari}
\thanks{Equal contributors}
\affiliation{Soft Matter Lab, Department of Physics, Bilkent University, Ankara 06800, Turkey}

\author{Yazgan Tuna}
\thanks{Equal contributors}
\affiliation{Soft Matter Lab, Department of Physics, Bilkent University, Ankara 06800, Turkey}
\affiliation{UNAM -- National Nanotechnology Research Center, Bilkent University, Ankara 06800, Turkey}

\author{Lukas Barth}
\affiliation{Soft Matter Lab, Department of Physics, Bilkent University, Ankara 06800, Turkey}

\author{Siegfried Dietrich}
\affiliation{Max-Planck-Institut f\"ur Intelligente Systeme, Heisenbergstr. 3, D-70569 Stuttgart, Germany, EU}
\affiliation{IV. Institut f\"ur Theoretische Physik, Universit\"at Stuttgart, Pfaffenwaldring 57, D-70569 Stuttgart, Germany, EU}

\author{Andrea Gambassi}
\affiliation{SISSA~--~International School for Advanced Studies and INFN, via Bonomea 265, 34136 Trieste, Italy, EU}

\author{Giovanni Volpe}
\email{giovanni.volpe@fen.bilkent.edu.tr}
\affiliation{Soft Matter Lab, Department of Physics, Bilkent University, Ankara 06800, Turkey}
\affiliation{UNAM -- National Nanotechnology Research Center, Bilkent University, Ankara 06800, Turkey}

\date{\today}

\begin{abstract}
In soft and condensed matter physics, effective interactions often emerge as a result of the spatial confinement of a fluctuating field. For instance, microscopic particles in a binary liquid mixture are subject to critical Casimir forces whenever their surfaces confine the thermal fluctuations of the order parameter of this kind of solvent, the range of which diverges upon approaching the critical demixing point. Critical Casimir forces are predicted to be nonadditive on a particular large scale. However, a direct experimental evidence of this fact is still lacking. Here, we fill in this gap by reporting the experimental measurement of the associated many-body effects. In particular, we focus on three colloidal particles in optical traps and observe that the critical Casimir force exerted on one of them by the other two colloids differs from the sum of the forces they exert separately. The magnitude and range of this three-body effect turn out to depend sensitively on the distance from the critical point of the solvent and on the preferential adsorption at the surfaces of the colloids for the two components of the mixture.
\end{abstract}

\maketitle


\textbf{Introduction.} From gravitation to electromagnetism, the fundamental physical interactions are additive: for instance, the force exerted on a probe electric charge in an homogeneous medium by two other charges equals the sum of the forces exerted by each of them separately. In more complex situations, however, effective interactions take hold --- and these forces are not necessarily additive \cite{israel1992}. Accordingly, a crucial issue is whether, and to what extent, genuinely many-body effects are present. For example, from the confinement of a fluctuating field, effective long-ranged forces between two microscopic objects may result at the micrometer scale. A notable example are critical Casimir forces acting on micrometer-sized particles which are immersed in a binary liquid mixture near its critical (demixing) point confining the concentration fluctuations \cite{fisher1978phenomena,gambassi2009rev}. These forces are theoretically predicted to be nonadditive \cite{mattos2013many,mattos2015manybody,hobrecht2015manybody}, but experimental evidence for the corresponding many-body effects is still lacking. Here, we report the direct experimental measurement of such effects. Using holographic optical tweezers (HOTs) \cite{grier2003revolution} and digital video microscopy (DVM) \cite{crocker1996methods,baumgartl2005limits} to probe \emph{in situ} the forces acting on spherical colloids immersed in a critical mixture of water and 2,6-lutidine in various geometrical configurations, we observe that the critical Casimir force exerted on a probe colloid  by two other colloids differs from the sum of the forces exerted by them separately. These many-body effects are controlled by adjusting the criticality of the mixture, e.g. by tuning its temperature or by changing the surface properties of the particles. Since interactions between microscopic particles in fluids are central to a wide spectrum of physical, chemical and biological phenomena, the insight we provide here might be useful in a diverse range of applications, including controlling microscopic self-assembly of colloids, complex collective behaviour, as well as phase and bio-mimetic behaviour of micro- and nanoparticles --- considering in particular that many-body effects are expected to become even more important on the nanoscale \cite{batista2015nonadditivity}.

Understanding and controlling  the interactions of micro- and nanoscopic particles is essential for a wide range of disciplines and applications, e.g. synthesis of complex nanodevices, self-organisation processes, and engineering of collective properties of nanoparticles. In these contexts, effective forces and the associated many-body effects can play a major role in determining large-scale collective and self-organised behaviours, e.g. concerning the formation and stabilisation of nanoparticle suspensions, aggregates, colloidal molecules and photonic crystals \cite{sacanna2010lock,batista2015nonadditivity}.

Effective long-ranged forces acting on mesoscopic objects emerge if they spatially confine a surrounding fluctuating field. QED Casimir forces are a notable example, due to the confinement of vacuum electromagnetic fluctuations between two conductors \cite{casimir1948attraction}; these forces are typically attractive and, thus, cause  undesired stiction of the metallic parts of nanodevices \cite{capasso2007casimir}. The thermodynamic analog of QED Casimir forces are critical Casimir forces, which were theoretically predicted by Fisher and de Gennes in 1978 \cite{fisher1978phenomena}: the confinement of thermal fluctuations in a binary liquid mixture may result in attractive or repulsive interactions, with universal features \cite{gambassi2009critical}. These thermal fluctuations typically occur on the molecular scale; however, upon approaching a critical point of a second-order phase transition, they become relevant and correlated across a much larger (up to several microns) length scale $\xi$. The first direct experimental evidence for critical Casimir forces was provided only in 2008 \cite{hertlein2008direct}: using total internal reflection microscopy (TIRM), femtonewton effective forces were experimentally measured between a single colloid and a planar surface immersed in a critical mixture; remarkably, both attractive and repulsive forces were found, in excellent agreement with theoretical predictions. Since then, these forces have been investigated under various conditions, e.g. by varying the properties of the involved surfaces \cite{nellen2009tunability,gambassi2011patterns}. In addition, a number of studies of the phase behaviour of colloidal dispersions in a critical mixture \cite{soyka2008critical,bonn2009direct,zvyagolskaya2011phase,mohry2012phase,edison2015} indicate critical Casimir forces as candidates for tuning the self-assembly of nanostructures \cite{faber2013controlling}. In order to gain full control and possibly harness critical Casimir interactions, it is pivotal to understand to what degree many-body effects play a role. In fact, both QED and critical Casimir forces have been theoretically predicted to be nonadditive \cite{mattos2013many,mattos2015manybody,hobrecht2015manybody}; however, a direct experimental evidence to support these theoretical predictions is still lacking.

Here we report a set of experiments which demonstrates and quantifies the three-body effects present in critical Casimir forces acting on colloidal microspheres immersed in a near-critical binary liquid mixture. Corresponding theoretical studies \cite{mattos2013many,mattos2015manybody,hobrecht2015manybody} reveal that these effects can either increase or decrease the critical Casimir forces depending on the temperature $T$ of the mixture, the spatial dimensionality, geometrical arrangement, shape, and distance between the involved surfaces, in a way that is difficult to rationalise but with an overall contribution which can be up to 20\% of the pairwise additive interaction. Due to this rather complex dependence on a large number of geometrical and physical variables, it is a priori unclear whether this effect can be experimentally detected in colloidal suspensions.


\medskip

\textbf{Experimental protocol.} Before getting into the experimental details, we briefly outline the strategy of the experiment. We employ a setup in which three colloidal microspheres are held by HOTs at the corners of an almost equilateral triangle. First, we measure the two-body critical Casimir forces arising, upon approaching criticality, on each of the three pairs of particles in the absence of the remaining colloid, which is temporarily moved into an auxiliary trap. Then, assuming additivity of critical Casimir forces, these measurements are used in order to predict the forces acting on a pair of particles in the presence of the third one, which is eventually brought closer. Finally, we compare this additive prediction with the actually measured three-body potential: the significant discrepancies which appear clearly demonstrate the nonadditivity of critical Casimir forces.

As solvent, we employ a mixture of water and 2,6-lutidine at the critical lutidine mass fraction $c_{\rm L}^{\rm c} = 0.286$ (see the phase diagram in Fig.~1a describing a lower critical point at temperature $T_{\rm c} = 306.4\,{\rm K}$ \cite{grattoni1993lower}). A few degrees below $T_{\rm c}$, the mixture is homogeneous and critical Casimir forces are negligible. However, as $T$ approaches $T_{\rm c}$ (arrow in Fig.~1a), critical concentration fluctuations emerge, which generate critical Casimir forces \cite{gambassi2009critical}. These forces depend strongly on the adsorption preferences of the surface of the particles, i.e. on whether they preferentially adsorb water or lutidine, realizing ($-$) and ($+$)  boundary conditions, respectively \cite{hertlein2008direct,gambassi2009critical}.  In particular, critical Casimir forces are attractive between two particles carrying the same boundary conditions, i.e. ($++$) and ($--$), while repulsive for ($+-$) and ($-+$). In the experiment
  we employ both hydrophilic ($-$) pristine silica spheres (diameter $2R = 2.06 \pm 0.05\,\mu$m) and hydrophobic ($+$) silica spheres obtained by treating their surfaces with octyltriemthoxysilane.

We start by considering the configuration schematically shown in Fig.~1b, with two particles, labeled by 1 and 2. These particles are subject to a one-body potential $U_{\rm ot}$ due to optical traps, and a two-body contribution due to both a screened electrostatic pair repulsion $U_{\rm es}$ and a  critical Casimir pair potential $U_{\rm C}$, while no evidence of van der Waals interactions was found;  the total potential $U_2$ of this configuration is therefore
\begin{equation}\label{eq:U2}
\begin{split}
U_2({\bf r}_1,{\bf r}_2) &\, = U^{(1)}_{\rm ot}(\Delta{\bf r}_1) + U^{(2)}_{\rm ot}(\Delta{\bf r}_2) \\
&\quad + U^{(12)}_{\rm es}(d_{12}) + U^{(12)}_{\rm C}(d_{12}, \xi)\,,
\end{split}
\end{equation}
where ${\bf r}_i$ is the position of the centre of particle $i$, $\Delta{\bf r}_i={\bf r}_i-{\bf R}_i$ and ${\bf R}_i$ is the position of the centre of the $i$-th trap, while $d_{ij} = |{\bf r}_i-{\bf r}_j|-2R$ is the distance between the surfaces of particles $i$ and $j$. The optical traps are harmonic \cite{jones2015optical}, i.e.
\begin{equation}\label{eq:ot}
U^{(i)}_{\rm ot}(\Delta{\bf r}_i) = \frac{k_i}{2}|\Delta{\bf r}_i|^2 \, ,
\end{equation}
where the stiffness $k_i$ ($\simeq 0.4\, \mbox{pN}/\mu\mbox{m}$ for the data in Figs.~1 and 3, and $\simeq 0.8\, \mbox{pN}/\mu\mbox{m}$ for those in Fig.~4) is nearly the same for all traps, as discussed below. The DLVO electrostatic potential \cite{israel1992} can be parameterised as
\begin{equation}\label{eq:dlvo}
U^{(ij)}_{\rm es}(d) = k_{\rm B} T \; \exp\left( - \frac{d - l^{(ij)}_{\rm es}}{l_{\rm D}} \right) \, ,
\end{equation}
where $k_{\rm B}$ is the Boltzmann constant, $l_{\rm D}$ ($\simeq 10\,{\rm nm}$) is the Debye length for the solvent, and $l^{(ij)}_{\rm es}$ ($\simeq 90\,{\rm nm}$) is a measure of the strength of the interaction between the two colloidal particles $i$ and $j$, which depends, inter alia, on their charges. The mild dependence of both $U_{\rm ot}$ and $U_{\rm es}$ on the temperature $T$ can be neglected  as
here $T$ is varied by at most 1\%. Finally, the critical Casimir pair potential $U^{(ij)}_{\rm C}(d_{ij},\xi)$ depends on the correlation length $\xi$ of the critical mixture and, within the Derjaguin approximation $d_{ij} \ll R$
(with $d_{ij}/R \lesssim 0.3$ in the present experiment), is given by \cite{gambassi2009critical,hertlein2008direct}
\begin{equation}\label{eq:C2}
U^{(ij)}_{\rm C}(d_{ij}, \xi) = k_{\rm B} T  \; \frac{R}{2 d} \Theta^{(ij)}(d_{ij}/\xi) \, ,
\end{equation}
where $\Theta^{(ij)}(x)$ is the universal scaling function characterised by the boundary conditions involved, but otherwise independent of the material properties of the mixture and of the particles; it can be inferred from available numerical data \cite{vasilyev2007monte,hertlein2008direct,gambassi2009critical}. The correlation length $\xi$ varies as $\xi(T) = \xi_0 ( 1- T/T_{\rm c})^{-\nu}$, where the mixture-specific quantity $\xi_0 = 0.20 \pm 0.02\,{\rm nm}$ (s.e.m.) has been determined by light scattering experiments \cite{gulari1972light}, while $\nu = 0.63$ is a universal critical exponent of the three-dimensional Ising universality class, holding for classical binary mixtures \cite{gambassi2009critical}. In the presence of three particles, their total potential $U_3$ can be decomposed into the sum of individual one-body, pairwise two-body, and remaining three-body contributions:
\begin{equation}\label{eq:U3}
\begin{split}
U_3({\bf r}_1,{\bf r}_2,{\bf r}_3)
&\;= \sum_{i=1}^{3} U^{(i)}_{\rm ot}({\bf r}_i-{\bf R}_i)  \\
&\ + \sum_{i, j=1,\, i<j}^{3}
\left[ U^{(ij)}_{\rm es}(d_{ij}) + U^{(ij)}_{\rm C}(d_{ij}, \xi)\right] \\[2mm]
&\ + U^{(123)}(d_{12},d_{13},d_{23}, \xi) \, ,
\end{split}
\end{equation}
where $U^{(ij)}_{\rm C}$ depends on the boundary conditions at the surfaces of the particles $i$ and $j$, while $U^{(123)}$ is the nonadditive three-body potential, which includes the contribution $U^{(123)}_{\rm C}$ of the three-body critical Casimir potential, the existence and magnitude of which we want to assess. Note that $U^{(123)}_{\rm C}$ can be distinguished from other possible contributions (such as those due to electrostatics \cite{brunner2004}) because of its sensitive dependence on temperature, which effectively controls also its spatial range.

Until now direct measurements of critical Casimir forces have been performed only on single particles above planar surfaces using TIRM \cite{hertlein2008direct,nellen2009tunability}; TIRM, however, necessarily involves a planar surface and therefore does not allow the measurement of forces arising between identical spherical particles, as they occur in colloidal suspensions. We have therefore developed an experimental setup capable of manipulating and observing multiple particles in the bulk of a critical mixture. Our setup is based on a combination of HOTs \cite{grier2003revolution,jones2015optical} and DVM \cite{crocker1996methods,baumgartl2005limits,jones2015optical}; its schematic is shown in Supplementary Fig.~1. The HOTs are realised by shaping a laser beam using a spatial light modulator; in this way, we generate multiple reconfigurable optical traps within the sample, which allow us to gently hold the colloids with nanometric accuracy in the bulk of the critical mixture, i.e. $50\,\mu \mbox{m}$ above the lower surface of the sample cell; this distance is large enough to ensure that the critical Casimir forces between this surface and the particles are negligible. The laser power at each trapping site is kept low enough ($\simeq 2\,{\rm mW}$) to avoid significant heating ($\ll 0.1\,{\rm mK}$ \cite{jones2015optical}). The DVM uses a monochromatic CCD camera, acquiring videos at $200\,{\rm fps}$; these frames are analysed using standard DVM algorithms in order to determine the projected positions of the particles with nanometric accuracy \cite{crocker1996methods,baumgartl2005limits}. The crucial temperature control of the binary mixture is achieved in two stages as described in the Methods and is able to maintain a certain temperature with an accuracy of $\pm 2\,{\rm mK}$.

We perform all measurements using the same configuration of six traps, all obtained by means of the  same hologram on the spatial light modulator, which produces spatially displaced but otherwise almost identical optical potentials. Three of these traps, referred to as internal ones, have their centres located at the vertices of an approximately equilateral triangle with edges of $\simeq 2.3\,\mu\mbox{m}$; the centres of the remaining three traps, referred to as external ones, are instead located at a distance $\simeq 2.3\,\mu\mbox{m}$ from each vertex, along the bisector of the corresponding external angle (see Fig.~2). The six traps are sufficiently far apart to guarantee the independence of the optical potentials \cite{jones2015optical}. The internal traps are used to hold either two (Figs.~2a, 2b and 2c) or three particles (Fig.~2d) for measuring the pair interaction or three-body potentials, respectively. The particles that are not needed in a certain measurement are temporarily moved from the internal to the closest of the external traps where they have negligible interactions with the other particles. In this way, the three colloids --- labeled 1, 2 and 3 --- selected at the beginning of the experiment  are employed for the whole set of measurements; this eliminates possible systematic errors due to  differences in the properties of the colloids of the batch we use. We start at a temperature $T$ a few degrees below $T_{\rm c}$, for which no critical fluctuations are present and the mixture is homogenous. First, we characterise the optical potential in Eq.~(\ref{eq:ot}) by measuring $k_i$ of each internal trap employing the configurations in Supplementary Fig.~2a: the particle under measurement (e.g. particle 1 in configuration a1) is held in an internal trap, while the other two particles (e.g. particles 2 and 3 in configuration a1) are moved into the nearest external traps. Then, by holding each pair (i.e. 1-2, 1-3 and 2-3) of particles in the respective internal traps while keeping the third particle (i.e. 3, 2 and 1, respectively) in the corresponding external trap, we measure and characterise the electrostatic interaction in Eq.~(\ref{eq:dlvo}) between each possible pair of particles, using the configurations shown in Figs.~2a, 2b and 2c, respectively. Finally, we measure the three-body interactions by  having all three colloids in the internal traps as shown in Fig.~2d. At temperatures much lower than $T_{\rm c}$, we obtain additivity of the interactions (in particular, of the electrostatic one) as expected. We then increase $T$ in small steps towards $T_{\rm c}$ and we repeat the measurement of the pair and three-body interactions at each step. The details of the measurement cycle are shown in Supplementary Fig.~2b.


\medskip

\textbf{Experimental results.} For each value of $T$, we acquire the histogram of the probability distribution $P_2(l_{12})$ of the in-plane surface-to-surface distance $l_{12}$ between the particles 1 and 2 (see Supplementary Fig.~3a and Methods for the definition of $l_{12}$) from which we infer the effective potential $U_2(l_{12}) \equiv - k_{\rm B}T\ln P_2(l_{12})$ shown by the symbols in Fig.~1c, for various values of the correlation length $\xi$ determined as described below. The solid lines in Fig.~1c represent, instead, the theoretical predictions based on Eq.~(\ref{eq:U2}) obtained via a Monte Carlo integration of the Boltzmann factor $\exp(-U_2({\bf r}_1,{\bf r}_2)/(k_{\rm B}T))$ (see the Methods for additional details) where the only fitting parameter is $\xi$, because the optical and the electrostatic potentials have already been characterised at low $T$ (see above). Note that the indicated values of $\xi$ have been determined by a best fit to a part of the experimental data (highlighted by darker colours in the histograms in Supplementary Figs.~3) and are the same for the three pairs of particles as $\xi$ only depends on the temperature of the mixture. We obtain a very good agreement between the measured and the theoretical effective potential for all three pairs of particles (Supplementary Figs.~3a, 3b and 3c). This demonstrates that Eq.~(\ref{eq:C2}) properly describes the two-body critical Casimir interaction $U_{\rm C}^{(12)}$, which is responsible for the formation of the dip in $U_2(l_{12})$ at $l_{12}\simeq 80\,\mbox{nm}$ and which appears as $\xi$ increases. In addition, this agreement provides the direct experimental evidence of the occurrence of critical Casimir forces between two spherical colloidal particles, which is a geometrical configuration which had not been previously explored. Note that $U_2(l_{12})$ at separations $l_{12} \gtrsim 250\,\mbox{nm}$ is essentially determined by $U_{\rm ot}$, while for $l_{12} \lesssim 50\,\mbox{nm}$ it is strongly influenced by the short-distance behaviour of $U_{\rm es}$ which might not be accurately captured by Eq.~(\ref{eq:dlvo}). Accordingly, the range of $l_{12}$ relevant for assessing the comparison between theory and experiment and the emergence of (possibly many-body) critical Casimir forces extends from the bottom of the dip to its right.

At this point, we can predict the effective potential $U_3(l_{12})$ associated with the distribution $P_3(l_{12})$ of the in-plane distance between particles 1 and 2 in the presence of particle 3 (Fig.~2d) by using the measured two-body interactions and assuming additivity, i.e. $U^{\rm (123)}\equiv 0$ in Eq.~(\ref{eq:U3}). Again, these theoretical predictions are computed numerically via a Monte Carlo integration of the Boltzmann factor $\exp(-U_3({\bf r}_1,{\bf r}_2, {\bf r}_3)/(k_{\rm B}T))$ (see the Methods for additional details). The resulting effective potential $U_3(l_{12})$ between particles 1 and 2 is indicated by the solid lines in Fig.~3, while symbols are the corresponding experimental data: the clear discrepancy between the two provides evidence for the presence of nonadditive (many-body) effects. The data with the smallest $\xi$ departs appreciably from additivity at short distances $l_{12}\lesssim 70\,\mbox{nm}$ due to a short-ranged electrostatic three-body effect which reduces repulsion \cite{brunner2004} and therefore amplifies the effects of the two-body critical Casimir attraction. On the contrary, as $\xi$ increases, the effective potential is less attractive than expected, indicating that such a reduction is due to the many-body critical Casimir interaction $U^{(123)}_{\rm C}$, especially for $l_{12}$ between 70 and 250 nm, where the two- and three-body electrostatics effects are negligible.

As pointed out above, one of the most distinguished features  of the two-body critical Casimir forces is that their attractive or repulsive character depends on the surface properties only of the particles. While the experiment described above involves three hydrophilic particles ($---$), we repeated the experiment with one hydrophobic (3) and two hydrophilic (1, 2) particles ($--+$), with the results shown in Fig.~4 and Supplementary Fig.~4. Also in this case we observe good agreement between theory and experiment for the pair-interaction effective potentials $U_2(l_{ij})$ which can be inferred from the histograms of the corresponding distribution $P_{2}(l_{ij})$ (see Supplementary Figs.~4a, 4b and 4c), while sizeable discrepancies emerge in the  three-body potential $U_3(l_{12})$ (Fig.~4 and Supplementary Fig.~4d). In contrast to the previous case, the experimental data demonstrate that, depending on the distance and the correlation length, many-body effects may also deepen the critical Casimir potential between particles 1 and 2. The experiments described above typically correspond to values of the scaling variables $R/\xi \simeq 100$ and $l_{ij}/R\simeq 0.1$ which are outside the ranges of the available theoretical predictions for the many-body effects \cite{mattos2015manybody}: their highly nontrivial dependence on temperature and geometrical features renders any attempt questionable to extrapolate  and therefore to compare these predictions with the present experimental data. Interestingly enough, assuming for $U^{(123)}_{\rm C}$ the simple functional form of the Axilrod-Teller three-atom potential which describes three-body corrections to the van der Waals interaction (see, e.g. Ref.~\cite{mattos2015manybody}), with a suitable choice of the overall amplitude, reduces the discrepancy between the experimental and the corresponding theoretical predictions.\\[3mm]

\smallskip

\textbf{Conclusions.} Our results provide direct evidence of the emergence of critical Casimir forces between two colloids and demonstrate the presence of pronounced three-body effects. These many-body critical Casimir forces strongly depend on the proximity to criticality of the fluid solvent, and can therefore be tuned, e.g. by changing its temperature or by altering the surface properties of the involved colloids. Criticality amounts to the occurrence of order parameter fluctuations on the spatial scale of the correlation length, which in principle can diverge and, thus, leads to the emergence of complex and nonadditive interactions at very large scales. They may find natural applications in various disciplines, such as in the realisation of colloidal molecules or reversible self-assembly, as well as the organisation of cellular membranes \cite{machta2012critical} and possibly even the patterns of brain activation \cite{haimovici2013brain}. Furthermore, in view of their similarity, these effects can also shine light on certain aspects of many-body effects concerning QED Casimir forces. While we focussed on configurations with three particles, the experimental setup and protocol discussed here are actually versatile enough to allow the investigation of the many-body potentials associated with a larger number of particles.

\smallskip

\textbf{Funding.} GV was partially funded by a Marie Curie Career Integration Grant (PCIG11GA-2012-321726) and a Distinguished Young Scientist award of the Turkish Academy of Sciences (T\"UBA).  This work was partially supported by Tubitak Grant No. 111T758. LB was supported by an internship awarded by DAAD Rise.

\section*{Methods}

\textbf{Experimental setup.} The experimental setup combines HOTs and DVM (Supplementary Figure~1) and is build around a homemade microscope. A laser beam (wavelength $532\,{\rm nm}$, power $500\,{\rm mW}$) is expanded by a telescope and projected onto a phase-only spatial light modulator (Holoeye, PLUTO-VIS). The hologram on the spatial light modulator imposes a phase modulation onto the beam, allowing the generation of multiple trapping spots. The resulting beam is then projected onto the entrance pupil of a high-numerical-aperture oil-immersion objective (magnification $100\times$, numerical aperture $1.30$) by a series of lenses and mirrors arranged in a 4f-configuration \cite{jones2015optical}. The objective focuses the beam in the sample plane and creates a series of reconfigurable traps. The DVM is realised using a standard configuration with white light illumination and a monochromatic CCD camera ($200\,{\rm fps}$). Since near $T_{\rm c}$ critical fluctuations depend very sensitively on small temperature changes, the microscope is enclosed within a thermally stabilised box in order to avoid any air flow, which may cause instability of the sample temperature. The sample holder is thermally isolated from the underneath translation stage with a teflon film. The necessary fine control of the temperature of the sample is achieved in two stages: first, the temperature of the microscope box is controlled to within $\pm 50\,{\rm mK}$; second, a closed-loop controller (realised with a Pt100 temperature sensor and a Peltier heating/cooling element) keeps the temperature of the sample cell to within $\pm 2\,{\rm mK}$. We remark that within each set of experiments the temperature $T$ is gradually increased towards $T_{\rm c}$.

\textbf{Sample preparation.} The binary liquid mixture used in all experiments is composed of water and 2,6-lutidine at the critical lutidine mass fraction $c_{\rm L}^{\rm c} = 0.286$. The mixture undergoes a second-order phase transition at a lower critical point with temperature $T_{\rm c} \cong 306.4\,{\rm K}$ \cite{grattoni1993lower}. The corresponding phase diagram is shown in Fig.~1a \cite{grattoni1993lower}: it consists of two regions, corresponding to the mixed (white) and the demixed (grey) states with a lower critical point (CP). In all experiments we use silica colloids with diameter $2R=2.06 \pm 0.05 \mu\mbox{m}$ (Microparticles GmbH). In the measurements involving hydrophobic particles, these silica colloids are treated chemically in order to make them hydrophobic, i.e. we silanise their surfaces with octyltriemthoxysilane. The sample cell containing the critical mixture and a small amount of colloids is a 200-$\mu$m-thick silica cuvette, which is sealed with teflon plugs in order to avoid evaporation of the mixture and to allow its usage for the whole duration of the experiment (ca.~1 day).

\textbf{Data analysis.} The raw data obtained from the various measurements are videos showing the orthogonal projection on a plane of the Brownian motion of the particles in three spatial dimensions and consisting of $\approx60\,000$~frames acquired at $200\,{\rm fps}$. From these videos we extract the in-plane position $(x_i,y_i)$ of the centre of each colloid $i$ as a function of time using standard DVM algorithms \cite{crocker1996methods,jones2015optical}, taking into account the necessary optical correction when the particles approach each other \cite{baumgartl2005limits}. Based on these trajectories we calculate the in-plane surface-to-surface distance $l_{ij} \equiv \sqrt{(x_i-x_j)^2 + (y_i-y_j)^2} - 2 R$ between particles $i$ and $j$, which allows us to construct the histograms $P_{2(3)}(l_{ij})$ corresponding to the acquired frames with two (three) close particles, and from them to determine the effective potentials $U_{2(3)}(l_{ij}) = - k_{\rm B}T\ln P_{2(3)}(l_{ij})$.

\textbf{Monte Carlo integration.} For each set of parameters we compute the theoretical in-plane surface-to-surface separation histograms $P_{2(3)}(l_{ij})$ and the associated effective potentials $U_{2(3)}(l_{ij})$ via a suitable Monte Carlo integration of the Boltzmann factors  $\exp(-U_2({\bf r}_1, {\bf r}_2)/(k_{\rm B}T))$ and $\exp(-U_3({\bf r}_1, {\bf r}_2, {\bf r}_3)/(k_{\rm B}T))$ with the theoretical potentials given by Eqs.~(\ref{eq:U2}), (\ref{eq:C2}) and (\ref{eq:U3}) with $U^{(123)}=0$ for the configurations with two and three close particles, respectively, based on the measured parameters for the optical traps and for the electrostatic interaction (Eq.~(\ref{eq:dlvo})). For instance, in the presence of three close particles, one has $P_3(l_{12}) \propto  \int \left(\prod_{i=1}^3 {\rm d}x_i {\rm d}y_i{\rm d}z_i \right)\delta(\sqrt{(x_1-x_2)^2 + (y_1-y_2)^2} - 2R - l_{12}) \exp(-U_3({\bf r}_1, {\bf r}_2, {\bf r}_3)/(k_{\rm B}T))$, where ${\bf r}_i \equiv (x_i,y_i,z_i)$ is the position of particle $i$ in a Cartesian coordinate system. In order to account for a (small) anisotropy of the generated optical traps, the data analysis and the fit are carried out by assuming  $U^{(i)}_{\rm ot}(\Delta{\bf r}) = \sum_{j=x,y,z} k_{i,j} ({\bf e}_j\cdot\Delta {\bf r})^2/2$, where ${\bf e}_{x,y,z}$ are the unit vectors along the principal orthogonal axes of the ellipsoidal trap, instead of Eq.~(\ref{eq:ot}). For all traps, ${\bf e}_{x,y}$ turn out to lie almost within the $x$--$y$ plane, with $k_{i,x} \simeq k_{i,y}$ and $k_{i,z} \simeq 0.3 k_{i,x}$. In particular, we verified that the actual values of $k_{i,z}$ do not significantly affect the comparison between the theoretical predictions and the experimental data.

\bibliography{biblio}

\begin{thebibliography}{30}%
\makeatletter
\providecommand \@ifxundefined [1]{%
 \@ifx{#1\undefined}
}%
\providecommand \@ifnum [1]{%
 \ifnum #1\expandafter \@firstoftwo
 \else \expandafter \@secondoftwo
 \fi
}%
\providecommand \@ifx [1]{%
 \ifx #1\expandafter \@firstoftwo
 \else \expandafter \@secondoftwo
 \fi
}%
\providecommand \natexlab [1]{#1}%
\providecommand \enquote  [1]{``#1''}%
\providecommand \bibnamefont  [1]{#1}%
\providecommand \bibfnamefont [1]{#1}%
\providecommand \citenamefont [1]{#1}%
\providecommand \href@noop [0]{\@secondoftwo}%
\providecommand \href [0]{\begingroup \@sanitize@url \@href}%
\providecommand \@href[1]{\@@startlink{#1}\@@href}%
\providecommand \@@href[1]{\endgroup#1\@@endlink}%
\providecommand \@sanitize@url [0]{\catcode `\\12\catcode `\$12\catcode
  `\&12\catcode `\#12\catcode `\^12\catcode `\_12\catcode `\%12\relax}%
\providecommand \@@startlink[1]{}%
\providecommand \@@endlink[0]{}%
\providecommand \url  [0]{\begingroup\@sanitize@url \@url }%
\providecommand \@url [1]{\endgroup\@href {#1}{\urlprefix }}%
\providecommand \urlprefix  [0]{URL }%
\providecommand \Eprint [0]{\href }%
\providecommand \doibase [0]{http://dx.doi.org/}%
\providecommand \selectlanguage [0]{\@gobble}%
\providecommand \bibinfo  [0]{\@secondoftwo}%
\providecommand \bibfield  [0]{\@secondoftwo}%
\providecommand \translation [1]{[#1]}%
\providecommand \BibitemOpen [0]{}%
\providecommand \bibitemStop [0]{}%
\providecommand \bibitemNoStop [0]{.\EOS\space}%
\providecommand \EOS [0]{\spacefactor3000\relax}%
\providecommand \BibitemShut  [1]{\csname bibitem#1\endcsname}%
\let\auto@bib@innerbib\@empty
\bibitem [{\citenamefont {Israelachvili}(1992)}]{israel1992}%
  \BibitemOpen
  \bibfield  {author} {\bibinfo {author} {\bibfnamefont {J.}~\bibnamefont
  {Israelachvili}},\ }\href@noop {} {\emph {\bibinfo {title} {Intermolecular
  and Surface Forces}}},\ \bibinfo {edition} {2nd}\ ed.\ (\bibinfo  {publisher}
  {Academic Press},\ \bibinfo {address} {London, UK},\ \bibinfo {year}
  {1992})\BibitemShut {NoStop}%
\bibitem [{\citenamefont {Fisher}\ and\ \citenamefont
  {de~Gennes}(1978)}]{fisher1978phenomena}%
  \BibitemOpen
  \bibfield  {author} {\bibinfo {author} {\bibfnamefont {M.~E.}\ \bibnamefont
  {Fisher}}\ and\ \bibinfo {author} {\bibfnamefont {P.~G.}\ \bibnamefont
  {de~Gennes}},\ }\bibfield  {title} {\enquote {\bibinfo {title} {Phenomena at
  the walls in a critical binary mixture},}\ }\href@noop {} {\bibfield
  {journal} {\bibinfo  {journal} {C. R. Acad. Sci. Paris B}\ }\textbf {\bibinfo
  {volume} {287}},\ \bibinfo {pages} {207--209} (\bibinfo {year}
  {1978})}\BibitemShut {NoStop}%
\bibitem [{\citenamefont {Gambassi}(2009)}]{gambassi2009rev}%
  \BibitemOpen
  \bibfield  {author} {\bibinfo {author} {\bibfnamefont {A.}~\bibnamefont
  {Gambassi}},\ }\bibfield  {title} {\enquote {\bibinfo {title} {The {C}asimir
  effect: From quantum to critical fluctuations},}\ }\href@noop {} {\bibfield
  {journal} {\bibinfo  {journal} {J. Phys.: Conf. Ser.}\ }\textbf {\bibinfo
  {volume} {161}},\ \bibinfo {pages} {012037} (\bibinfo {year}
  {2009})}\BibitemShut {NoStop}%
\bibitem [{\citenamefont {Mattos}\ \emph {et~al.}(2013)\citenamefont {Mattos},
  \citenamefont {Harnau},\ and\ \citenamefont {Dietrich}}]{mattos2013many}%
  \BibitemOpen
  \bibfield  {author} {\bibinfo {author} {\bibfnamefont {T.~G.}\ \bibnamefont
  {Mattos}}, \bibinfo {author} {\bibfnamefont {L.}~\bibnamefont {Harnau}}, \
  and\ \bibinfo {author} {\bibfnamefont {S.}~\bibnamefont {Dietrich}},\
  }\bibfield  {title} {\enquote {\bibinfo {title} {Many-body effects for
  critical {C}asimir forces},}\ }\href@noop {} {\bibfield  {journal} {\bibinfo
  {journal} {J. Chem. Phys.}\ }\textbf {\bibinfo {volume} {138}},\ \bibinfo
  {pages} {074704} (\bibinfo {year} {2013})}\BibitemShut {NoStop}%
\bibitem [{\citenamefont {Mattos}\ \emph {et~al.}(2015)\citenamefont {Mattos},
  \citenamefont {Harnau},\ and\ \citenamefont {Dietrich}}]{mattos2015manybody}%
  \BibitemOpen
  \bibfield  {author} {\bibinfo {author} {\bibfnamefont {T.~G.}\ \bibnamefont
  {Mattos}}, \bibinfo {author} {\bibfnamefont {L.}~\bibnamefont {Harnau}}, \
  and\ \bibinfo {author} {\bibfnamefont {S.}~\bibnamefont {Dietrich}},\
  }\bibfield  {title} {\enquote {\bibinfo {title} {Three-body critical
  {C}asimir forces},}\ }\href@noop {} {\bibfield  {journal} {\bibinfo
  {journal} {Phys. Rev. E}\ }\textbf {\bibinfo {volume} {91}},\ \bibinfo
  {pages} {042304} (\bibinfo {year} {2015})}\BibitemShut {NoStop}%
\bibitem [{\citenamefont {Hobrecht}\ and\ \citenamefont
  {Hucht}(2015)}]{hobrecht2015manybody}%
  \BibitemOpen
  \bibfield  {author} {\bibinfo {author} {\bibfnamefont {H.}~\bibnamefont
  {Hobrecht}}\ and\ \bibinfo {author} {\bibfnamefont {A.}~\bibnamefont
  {Hucht}},\ }\bibfield  {title} {\enquote {\bibinfo {title} {Many-body
  critical {C}asimir interactions in colloidal suspensions},}\ }\href@noop {}
  {\bibfield  {journal} {\bibinfo  {journal} {Phys. Rev. E}\ }\textbf {\bibinfo
  {volume} {92}},\ \bibinfo {pages} {042315} (\bibinfo {year}
  {2015})}\BibitemShut {NoStop}%
\bibitem [{\citenamefont {Grier}(2003)}]{grier2003revolution}%
  \BibitemOpen
  \bibfield  {author} {\bibinfo {author} {\bibfnamefont {D.~G}\ \bibnamefont
  {Grier}},\ }\bibfield  {title} {\enquote {\bibinfo {title} {A revolution in
  optical manipulation},}\ }\href@noop {} {\bibfield  {journal} {\bibinfo
  {journal} {Nature}\ }\textbf {\bibinfo {volume} {424}},\ \bibinfo {pages}
  {810--816} (\bibinfo {year} {2003})}\BibitemShut {NoStop}%
\bibitem [{\citenamefont {Crocker}\ and\ \citenamefont
  {Grier}(1996)}]{crocker1996methods}%
  \BibitemOpen
  \bibfield  {author} {\bibinfo {author} {\bibfnamefont {John~C}\ \bibnamefont
  {Crocker}}\ and\ \bibinfo {author} {\bibfnamefont {David~G}\ \bibnamefont
  {Grier}},\ }\bibfield  {title} {\enquote {\bibinfo {title} {Methods of
  digital video microscopy for colloidal studies},}\ }\href@noop {} {\bibfield
  {journal} {\bibinfo  {journal} {J. Colloid Interface Sci.}\ }\textbf
  {\bibinfo {volume} {179}},\ \bibinfo {pages} {298--310} (\bibinfo {year}
  {1996})}\BibitemShut {NoStop}%
\bibitem [{\citenamefont {Baumgartl}\ and\ \citenamefont
  {Bechinger}(2005)}]{baumgartl2005limits}%
  \BibitemOpen
  \bibfield  {author} {\bibinfo {author} {\bibfnamefont {J.}~\bibnamefont
  {Baumgartl}}\ and\ \bibinfo {author} {\bibfnamefont {C.}~\bibnamefont
  {Bechinger}},\ }\bibfield  {title} {\enquote {\bibinfo {title} {On the limits
  of digital video microscopy},}\ }\href@noop {} {\bibfield  {journal}
  {\bibinfo  {journal} {EPL (Europhys. Lett.)}\ }\textbf {\bibinfo {volume}
  {71}},\ \bibinfo {pages} {487} (\bibinfo {year} {2005})}\BibitemShut
  {NoStop}%
\bibitem [{\citenamefont {Batista}\ \emph {et~al.}(2015)\citenamefont
  {Batista}, \citenamefont {Larson},\ and\ \citenamefont
  {Kotov}}]{batista2015nonadditivity}%
  \BibitemOpen
  \bibfield  {author} {\bibinfo {author} {\bibfnamefont {C.~A.~S.}\
  \bibnamefont {Batista}}, \bibinfo {author} {\bibfnamefont {R.~G.}\
  \bibnamefont {Larson}}, \ and\ \bibinfo {author} {\bibfnamefont {N.~A.}\
  \bibnamefont {Kotov}},\ }\bibfield  {title} {\enquote {\bibinfo {title}
  {Nonadditivity of nanoparticle interactions},}\ }\href@noop {} {\bibfield
  {journal} {\bibinfo  {journal} {Science}\ }\textbf {\bibinfo {volume}
  {350}},\ \bibinfo {pages} {1242477} (\bibinfo {year} {2015})}\BibitemShut
  {NoStop}%
\bibitem [{\citenamefont {Sacanna}\ \emph {et~al.}(2010)\citenamefont
  {Sacanna}, \citenamefont {Irvine}, \citenamefont {Chaikin},\ and\
  \citenamefont {Pine}}]{sacanna2010lock}%
  \BibitemOpen
  \bibfield  {author} {\bibinfo {author} {\bibfnamefont {S.}~\bibnamefont
  {Sacanna}}, \bibinfo {author} {\bibfnamefont {W.~T.~M.}\ \bibnamefont
  {Irvine}}, \bibinfo {author} {\bibfnamefont {P.~M.}\ \bibnamefont {Chaikin}},
  \ and\ \bibinfo {author} {\bibfnamefont {D.~J.}\ \bibnamefont {Pine}},\
  }\bibfield  {title} {\enquote {\bibinfo {title} {Lock and key colloids},}\
  }\href@noop {} {\bibfield  {journal} {\bibinfo  {journal} {Nature}\ }\textbf
  {\bibinfo {volume} {464}},\ \bibinfo {pages} {575--578} (\bibinfo {year}
  {2010})}\BibitemShut {NoStop}%
\bibitem [{\citenamefont {Casimir}(1948)}]{casimir1948attraction}%
  \BibitemOpen
  \bibfield  {author} {\bibinfo {author} {\bibfnamefont {H.~B.~G.}\
  \bibnamefont {Casimir}},\ }\bibfield  {title} {\enquote {\bibinfo {title} {On
  the attraction between two perfectly conducting plates},}\ }\href@noop {}
  {\bibfield  {journal} {\bibinfo  {journal} {Proc. K. Ned. Akad. Wet}\
  }\textbf {\bibinfo {volume} {B51}},\ \bibinfo {pages} {793--795} (\bibinfo
  {year} {1948})}\BibitemShut {NoStop}%
\bibitem [{\citenamefont {Capasso}\ \emph {et~al.}(2007)\citenamefont
  {Capasso}, \citenamefont {Munday}, \citenamefont {Iannuzzi},\ and\
  \citenamefont {Chan}}]{capasso2007casimir}%
  \BibitemOpen
  \bibfield  {author} {\bibinfo {author} {\bibfnamefont {F.}~\bibnamefont
  {Capasso}}, \bibinfo {author} {\bibfnamefont {J.~N.}\ \bibnamefont {Munday}},
  \bibinfo {author} {\bibfnamefont {D.}~\bibnamefont {Iannuzzi}}, \ and\
  \bibinfo {author} {\bibfnamefont {H.~B.}\ \bibnamefont {Chan}},\ }\bibfield
  {title} {\enquote {\bibinfo {title} {{C}asimir forces and quantum
  electrodynamical torques: Physics and nanomechanics},}\ }\href@noop {}
  {\bibfield  {journal} {\bibinfo  {journal} {Sel. Top. Quant. El., IEEE J.}\
  }\textbf {\bibinfo {volume} {13}},\ \bibinfo {pages} {400--414} (\bibinfo
  {year} {2007})}\BibitemShut {NoStop}%
\bibitem [{\citenamefont {Gambassi}\ \emph {et~al.}(2009)\citenamefont
  {Gambassi}, \citenamefont {Macio{\l}ek}, \citenamefont {Hertlein},
  \citenamefont {Nellen}, \citenamefont {Helden}, \citenamefont {Bechinger},\
  and\ \citenamefont {Dietrich}}]{gambassi2009critical}%
  \BibitemOpen
  \bibfield  {author} {\bibinfo {author} {\bibfnamefont {A.}~\bibnamefont
  {Gambassi}}, \bibinfo {author} {\bibfnamefont {A.}~\bibnamefont
  {Macio{\l}ek}}, \bibinfo {author} {\bibfnamefont {C.}~\bibnamefont
  {Hertlein}}, \bibinfo {author} {\bibfnamefont {U.}~\bibnamefont {Nellen}},
  \bibinfo {author} {\bibfnamefont {L.}~\bibnamefont {Helden}}, \bibinfo
  {author} {\bibfnamefont {C.}~\bibnamefont {Bechinger}}, \ and\ \bibinfo
  {author} {\bibfnamefont {S.}~\bibnamefont {Dietrich}},\ }\bibfield  {title}
  {\enquote {\bibinfo {title} {Critical {C}asimir effect in classical binary
  liquid mixtures},}\ }\href@noop {} {\bibfield  {journal} {\bibinfo  {journal}
  {Phys. Rev. E}\ }\textbf {\bibinfo {volume} {80}},\ \bibinfo {pages} {061143}
  (\bibinfo {year} {2009})}\BibitemShut {NoStop}%
\bibitem [{\citenamefont {Hertlein}\ \emph {et~al.}(2008)\citenamefont
  {Hertlein}, \citenamefont {Helden}, \citenamefont {Gambassi}, \citenamefont
  {Dietrich},\ and\ \citenamefont {Bechinger}}]{hertlein2008direct}%
  \BibitemOpen
  \bibfield  {author} {\bibinfo {author} {\bibfnamefont {C.}~\bibnamefont
  {Hertlein}}, \bibinfo {author} {\bibfnamefont {L.}~\bibnamefont {Helden}},
  \bibinfo {author} {\bibfnamefont {A.}~\bibnamefont {Gambassi}}, \bibinfo
  {author} {\bibfnamefont {S.}~\bibnamefont {Dietrich}}, \ and\ \bibinfo
  {author} {\bibfnamefont {C.}~\bibnamefont {Bechinger}},\ }\bibfield  {title}
  {\enquote {\bibinfo {title} {Direct measurement of critical {C}asimir
  forces},}\ }\href@noop {} {\bibfield  {journal} {\bibinfo  {journal}
  {Nature}\ }\textbf {\bibinfo {volume} {451}},\ \bibinfo {pages} {172--175}
  (\bibinfo {year} {2008})}\BibitemShut {NoStop}%
\bibitem [{\citenamefont {Nellen}\ \emph {et~al.}(2009)\citenamefont {Nellen},
  \citenamefont {Helden},\ and\ \citenamefont
  {Bechinger}}]{nellen2009tunability}%
  \BibitemOpen
  \bibfield  {author} {\bibinfo {author} {\bibfnamefont {U.}~\bibnamefont
  {Nellen}}, \bibinfo {author} {\bibfnamefont {L.}~\bibnamefont {Helden}}, \
  and\ \bibinfo {author} {\bibfnamefont {C.}~\bibnamefont {Bechinger}},\
  }\bibfield  {title} {\enquote {\bibinfo {title} {Tunability of critical
  {C}asimir interactions by boundary conditions},}\ }\href@noop {} {\bibfield
  {journal} {\bibinfo  {journal} {EPL (Europhys. Lett.)}\ }\textbf {\bibinfo
  {volume} {88}},\ \bibinfo {pages} {26001} (\bibinfo {year}
  {2009})}\BibitemShut {NoStop}%
\bibitem [{\citenamefont {Gambassi}\ and\ \citenamefont
  {Dietrich}(2011)}]{gambassi2011patterns}%
  \BibitemOpen
  \bibfield  {author} {\bibinfo {author} {\bibfnamefont {A.}~\bibnamefont
  {Gambassi}}\ and\ \bibinfo {author} {\bibfnamefont {S.}~\bibnamefont
  {Dietrich}},\ }\bibfield  {title} {\enquote {\bibinfo {title} {Critical
  {C}asimir forces steered by patterned substrates},}\ }\href@noop {}
  {\bibfield  {journal} {\bibinfo  {journal} {Soft Matter}\ }\textbf {\bibinfo
  {volume} {7}},\ \bibinfo {pages} {1247} (\bibinfo {year} {2011})}\BibitemShut
  {NoStop}%
\bibitem [{\citenamefont {Soyka}\ \emph {et~al.}(2008)\citenamefont {Soyka},
  \citenamefont {Zvyagolskaya}, \citenamefont {Hertlein}, \citenamefont
  {Helden},\ and\ \citenamefont {Bechinger}}]{soyka2008critical}%
  \BibitemOpen
  \bibfield  {author} {\bibinfo {author} {\bibfnamefont {F.}~\bibnamefont
  {Soyka}}, \bibinfo {author} {\bibfnamefont {O.}~\bibnamefont {Zvyagolskaya}},
  \bibinfo {author} {\bibfnamefont {C.}~\bibnamefont {Hertlein}}, \bibinfo
  {author} {\bibfnamefont {L.}~\bibnamefont {Helden}}, \ and\ \bibinfo {author}
  {\bibfnamefont {C.}~\bibnamefont {Bechinger}},\ }\bibfield  {title} {\enquote
  {\bibinfo {title} {Critical {C}asimir forces in colloidal suspensions on
  chemically patterned surfaces},}\ }\href@noop {} {\bibfield  {journal}
  {\bibinfo  {journal} {Phys. Rev. Lett.}\ }\textbf {\bibinfo {volume} {101}},\
  \bibinfo {pages} {208301} (\bibinfo {year} {2008})}\BibitemShut {NoStop}%
\bibitem [{\citenamefont {Bonn}\ \emph {et~al.}(2009)\citenamefont {Bonn},
  \citenamefont {Otwinowski}, \citenamefont {Sacanna}, \citenamefont {Guo},
  \citenamefont {Wegdam},\ and\ \citenamefont {Schall}}]{bonn2009direct}%
  \BibitemOpen
  \bibfield  {author} {\bibinfo {author} {\bibfnamefont {D.}~\bibnamefont
  {Bonn}}, \bibinfo {author} {\bibfnamefont {J.}~\bibnamefont {Otwinowski}},
  \bibinfo {author} {\bibfnamefont {S.}~\bibnamefont {Sacanna}}, \bibinfo
  {author} {\bibfnamefont {H.}~\bibnamefont {Guo}}, \bibinfo {author}
  {\bibfnamefont {G.}~\bibnamefont {Wegdam}}, \ and\ \bibinfo {author}
  {\bibfnamefont {P.}~\bibnamefont {Schall}},\ }\bibfield  {title} {\enquote
  {\bibinfo {title} {Direct observation of colloidal aggregation by critical
  {C}asimir forces},}\ }\href@noop {} {\bibfield  {journal} {\bibinfo
  {journal} {Phys. Rev. Lett.}\ }\textbf {\bibinfo {volume} {103}},\ \bibinfo
  {pages} {156101} (\bibinfo {year} {2009})}\BibitemShut {NoStop}%
\bibitem [{\citenamefont {Zvyagolskaya}\ \emph {et~al.}(2011)\citenamefont
  {Zvyagolskaya}, \citenamefont {Archer},\ and\ \citenamefont
  {Bechinger}}]{zvyagolskaya2011phase}%
  \BibitemOpen
  \bibfield  {author} {\bibinfo {author} {\bibfnamefont {O.}~\bibnamefont
  {Zvyagolskaya}}, \bibinfo {author} {\bibfnamefont {A.~J.}\ \bibnamefont
  {Archer}}, \ and\ \bibinfo {author} {\bibfnamefont {C.}~\bibnamefont
  {Bechinger}},\ }\bibfield  {title} {\enquote {\bibinfo {title} {Criticality
  and phase separation in a two-dimensional binary colloidal fluid induced by
  the solvent critical behavior},}\ }\href@noop {} {\bibfield  {journal}
  {\bibinfo  {journal} {EPL (Europhys. Lett.)}\ }\textbf {\bibinfo {volume}
  {96}},\ \bibinfo {pages} {28005} (\bibinfo {year} {2011})}\BibitemShut
  {NoStop}%
\bibitem [{\citenamefont {Mohry}\ \emph {et~al.}(2012)\citenamefont {Mohry},
  \citenamefont {Macio{\l}ek},\ and\ \citenamefont
  {Dietrich}}]{mohry2012phase}%
  \BibitemOpen
  \bibfield  {author} {\bibinfo {author} {\bibfnamefont {T.~F.}\ \bibnamefont
  {Mohry}}, \bibinfo {author} {\bibfnamefont {A.}~\bibnamefont {Macio{\l}ek}},
  \ and\ \bibinfo {author} {\bibfnamefont {S.}~\bibnamefont {Dietrich}},\
  }\bibfield  {title} {\enquote {\bibinfo {title} {Phase behavior of colloidal
  suspensions with critical solvents in terms of effective interactions},}\
  }\href@noop {} {\bibfield  {journal} {\bibinfo  {journal} {J. Chem. Phys.}\
  }\textbf {\bibinfo {volume} {136}},\ \bibinfo {pages} {224902} (\bibinfo
  {year} {2012})}\BibitemShut {NoStop}%
\bibitem [{\citenamefont {Edison}\ \emph {et~al.}(2015)\citenamefont {Edison},
  \citenamefont {Tasios}, \citenamefont {Belli}, \citenamefont {Evans},
  \citenamefont {van Roij},\ and\ \citenamefont {Dijkstra}}]{edison2015}%
  \BibitemOpen
  \bibfield  {author} {\bibinfo {author} {\bibfnamefont {J.~R.}\ \bibnamefont
  {Edison}}, \bibinfo {author} {\bibfnamefont {N.}~\bibnamefont {Tasios}},
  \bibinfo {author} {\bibfnamefont {S.}~\bibnamefont {Belli}}, \bibinfo
  {author} {\bibfnamefont {R.}~\bibnamefont {Evans}}, \bibinfo {author}
  {\bibfnamefont {R.}~\bibnamefont {van Roij}}, \ and\ \bibinfo {author}
  {\bibfnamefont {M.}~\bibnamefont {Dijkstra}},\ }\bibfield  {title} {\enquote
  {\bibinfo {title} {Critical {C}asimir forces and colloidal phase transitions
  in a near-critical solvent: A simple model reveals a rich phase diagram},}\
  }\href@noop {} {\bibfield  {journal} {\bibinfo  {journal} {Phys. Rev. Lett.}\
  }\textbf {\bibinfo {volume} {114}},\ \bibinfo {pages} {038301} (\bibinfo
  {year} {2015})}\BibitemShut {NoStop}%
\bibitem [{\citenamefont {Faber}\ \emph {et~al.}(2013)\citenamefont {Faber},
  \citenamefont {Hu}, \citenamefont {Wegdam},\ and\ \citenamefont
  {Schall}}]{faber2013controlling}%
  \BibitemOpen
  \bibfield  {author} {\bibinfo {author} {\bibfnamefont {S.}~\bibnamefont
  {Faber}}, \bibinfo {author} {\bibfnamefont {Z.}~\bibnamefont {Hu}}, \bibinfo
  {author} {\bibfnamefont {G.~H.}\ \bibnamefont {Wegdam}}, \ and\ \bibinfo
  {author} {\bibfnamefont {P.}~\bibnamefont {Schall}},\ }\bibfield  {title}
  {\enquote {\bibinfo {title} {Controlling colloidal phase transitions with
  critical {C}asimir forces},}\ }\href@noop {} {\bibfield  {journal} {\bibinfo
  {journal} {Nature Commun.}\ }\textbf {\bibinfo {volume} {4}},\ \bibinfo
  {pages} {1584} (\bibinfo {year} {2013})}\BibitemShut {NoStop}%
\bibitem [{\citenamefont {Grattoni}\ \emph {et~al.}(1993)\citenamefont
  {Grattoni}, \citenamefont {Dawe}, \citenamefont {Seah},\ and\ \citenamefont
  {Gray}}]{grattoni1993lower}%
  \BibitemOpen
  \bibfield  {author} {\bibinfo {author} {\bibfnamefont {C.~A.}\ \bibnamefont
  {Grattoni}}, \bibinfo {author} {\bibfnamefont {R.~A.}\ \bibnamefont {Dawe}},
  \bibinfo {author} {\bibfnamefont {C.~Y.}\ \bibnamefont {Seah}}, \ and\
  \bibinfo {author} {\bibfnamefont {J.~D.}\ \bibnamefont {Gray}},\ }\bibfield
  {title} {\enquote {\bibinfo {title} {Lower critical solution coexistence
  curve and physical properties (density, viscosity, surface tension, and
  interfacial tension) of 2, 6-lutidine+water},}\ }\href@noop {} {\bibfield
  {journal} {\bibinfo  {journal} {J. Chem. Eng. Data}\ }\textbf {\bibinfo
  {volume} {38}},\ \bibinfo {pages} {516--519} (\bibinfo {year}
  {1993})}\BibitemShut {NoStop}%
\bibitem [{\citenamefont {Jones}\ \emph {et~al.}(2015)\citenamefont {Jones},
  \citenamefont {Marag\`o},\ and\ \citenamefont {Volpe}}]{jones2015optical}%
  \BibitemOpen
  \bibfield  {author} {\bibinfo {author} {\bibfnamefont {P.~H.}\ \bibnamefont
  {Jones}}, \bibinfo {author} {\bibfnamefont {O.~M.}\ \bibnamefont {Marag\`o}},
  \ and\ \bibinfo {author} {\bibfnamefont {G.}~\bibnamefont {Volpe}},\
  }\href@noop {} {\emph {\bibinfo {title} {Optical tweezers: Principles and
  applications}}}\ (\bibinfo  {publisher} {Cambridge University Press},\
  \bibinfo {address} {Cambridge, UK},\ \bibinfo {year} {2015})\BibitemShut
  {NoStop}%
\bibitem [{\citenamefont {Vasilyev}\ \emph {et~al.}(2007)\citenamefont
  {Vasilyev}, \citenamefont {Gambassi}, \citenamefont {Macio{\l}ek},\ and\
  \citenamefont {Dietrich}}]{vasilyev2007monte}%
  \BibitemOpen
  \bibfield  {author} {\bibinfo {author} {\bibfnamefont {O.}~\bibnamefont
  {Vasilyev}}, \bibinfo {author} {\bibfnamefont {A.}~\bibnamefont {Gambassi}},
  \bibinfo {author} {\bibfnamefont {A.}~\bibnamefont {Macio{\l}ek}}, \ and\
  \bibinfo {author} {\bibfnamefont {S.}~\bibnamefont {Dietrich}},\ }\bibfield
  {title} {\enquote {\bibinfo {title} {Monte {C}arlo simulation results for
  critical {C}asimir forces},}\ }\href@noop {} {\bibfield  {journal} {\bibinfo
  {journal} {EPL (Europhys. Lett.)}\ }\textbf {\bibinfo {volume} {80}},\
  \bibinfo {pages} {60009} (\bibinfo {year} {2007})}\BibitemShut {NoStop}%
\bibitem [{\citenamefont {G{\"u}lari}\ \emph {et~al.}(1972)\citenamefont
  {G{\"u}lari}, \citenamefont {Collings}, \citenamefont {Schmidt},\ and\
  \citenamefont {Pings}}]{gulari1972light}%
  \BibitemOpen
  \bibfield  {author} {\bibinfo {author} {\bibfnamefont {E.}~\bibnamefont
  {G{\"u}lari}}, \bibinfo {author} {\bibfnamefont {A.~F.}\ \bibnamefont
  {Collings}}, \bibinfo {author} {\bibfnamefont {R.~L.}\ \bibnamefont
  {Schmidt}}, \ and\ \bibinfo {author} {\bibfnamefont {C.~J.}\ \bibnamefont
  {Pings}},\ }\bibfield  {title} {\enquote {\bibinfo {title} {Light scattering
  and shear viscosity studies of the binary system 2,6-lutidine-water in the
  critical region},}\ }\href@noop {} {\bibfield  {journal} {\bibinfo  {journal}
  {J. Chem. Phys.}\ }\textbf {\bibinfo {volume} {56}},\ \bibinfo {pages}
  {6169--6179} (\bibinfo {year} {1972})}\BibitemShut {NoStop}%
\bibitem [{\citenamefont {Brunner}\ \emph {et~al.}(2004)\citenamefont
  {Brunner}, \citenamefont {Dobnikar}, \citenamefont {von Gr\"unberg},\ and\
  \citenamefont {Bechinger}}]{brunner2004}%
  \BibitemOpen
  \bibfield  {author} {\bibinfo {author} {\bibfnamefont {M.}~\bibnamefont
  {Brunner}}, \bibinfo {author} {\bibfnamefont {J.}~\bibnamefont {Dobnikar}},
  \bibinfo {author} {\bibfnamefont {H.-H.}\ \bibnamefont {von Gr\"unberg}}, \
  and\ \bibinfo {author} {\bibfnamefont {C.}~\bibnamefont {Bechinger}},\
  }\bibfield  {title} {\enquote {\bibinfo {title} {Direct measurement of
  three-body interactions amongst charged colloids},}\ }\href@noop {}
  {\bibfield  {journal} {\bibinfo  {journal} {Phys. Rev. Lett.}\ }\textbf
  {\bibinfo {volume} {92}},\ \bibinfo {pages} {078301} (\bibinfo {year}
  {2004})}\BibitemShut {NoStop}%
\bibitem [{\citenamefont {Machta}\ \emph {et~al.}(2012)\citenamefont {Machta},
  \citenamefont {Veatch},\ and\ \citenamefont {Sethna}}]{machta2012critical}%
  \BibitemOpen
  \bibfield  {author} {\bibinfo {author} {\bibfnamefont {B.~B.}\ \bibnamefont
  {Machta}}, \bibinfo {author} {\bibfnamefont {S.~L.}\ \bibnamefont {Veatch}},
  \ and\ \bibinfo {author} {\bibfnamefont {J.~P.}\ \bibnamefont {Sethna}},\
  }\bibfield  {title} {\enquote {\bibinfo {title} {Critical {C}asimir forces in
  cellular membranes},}\ }\href@noop {} {\bibfield  {journal} {\bibinfo
  {journal} {Phys. Rev. Lett.}\ }\textbf {\bibinfo {volume} {109}},\ \bibinfo
  {pages} {138101} (\bibinfo {year} {2012})}\BibitemShut {NoStop}%
\bibitem [{\citenamefont {Haimovici}\ \emph {et~al.}(2013)\citenamefont
  {Haimovici}, \citenamefont {Tagliazucchi}, \citenamefont {Balenzuela},\ and\
  \citenamefont {Chialvo}}]{haimovici2013brain}%
  \BibitemOpen
  \bibfield  {author} {\bibinfo {author} {\bibfnamefont {A.}~\bibnamefont
  {Haimovici}}, \bibinfo {author} {\bibfnamefont {E.}~\bibnamefont
  {Tagliazucchi}}, \bibinfo {author} {\bibfnamefont {P.}~\bibnamefont
  {Balenzuela}}, \ and\ \bibinfo {author} {\bibfnamefont {D.~R.}\ \bibnamefont
  {Chialvo}},\ }\bibfield  {title} {\enquote {\bibinfo {title} {Brain
  organization into resting state networks emerges at criticality on a model of
  the human connectome},}\ }\href@noop {} {\bibfield  {journal} {\bibinfo
  {journal} {Phys. Rev. Lett.}\ }\textbf {\bibinfo {volume} {110}},\ \bibinfo
  {pages} {178101} (\bibinfo {year} {2013})}\BibitemShut {NoStop}%
\end{thebibliography}%



%
%

\clearpage

\begin{widetext}

\renewcommand{\figurename}{{\bf Figure}}
\setcounter{figure}{0}
\renewcommand{\thefigure}{{\bf \arabic{figure}}}
\begin{figure}
\includegraphics[width=0.7\textwidth]{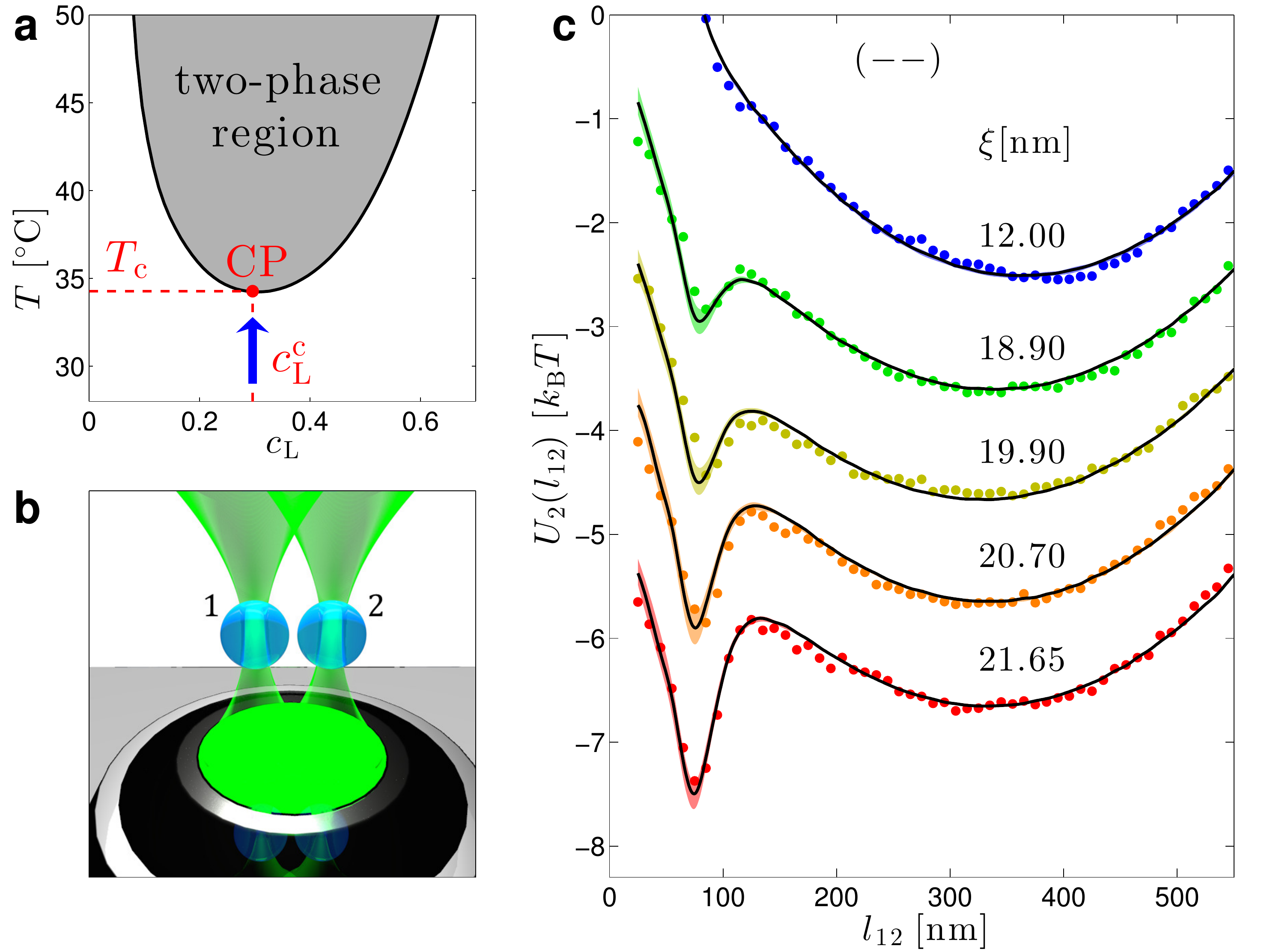}
\caption{
{\bf Measurement of the effective potentials between two optically trapped microspheres in a critical mixture.}
{\bf a}, Phase diagram of the water--2,6-lutidine mixture featuring a lower critical point (CP) at the bottom of the coexistence line (thick solid line \cite{grattoni1993lower}). Measurements are performed at the critical lutidine mass fraction $c_{\rm L}^{\rm c} = 0.286$, while the temperature $T$ is gradually increased towards its critical value $T_{\rm c} = 306.4\,{\rm K}$, as indicated by the arrow.
{\bf b}, Cartoon of the experimental setup for the measurement of effective pair interactions: two spherical silica colloids (blue spheres, diameter $2R = 2.06\pm 0.05\,{\rm \mu m}$) are held in the bulk of the binary mixture (not shown) by two optical tweezers (green conoids) obtained by focusing a laser beam via the objective indicated directly below (black and silver against a grey background). While the size of the particles is to scale with their relative distance, the objective and its distance from the particle are not.
{\bf c}, Effective pair potential $U_2(l_{12})$ between two hydrophilic colloids labelled 1 and 2 (with boundary conditions ($--$), i.e. attractive critical Casimir forces) as a function of the in-plane surface-to-surface distance $l_{12}$: the symbols represent the experimental data and the solid lines the theoretical fits with the associated uncertainty (shading). From top to bottom, $T$ increases towards $T_{\rm c}$, which is accompanied by an increase of the fitted correlation length $\xi$ and by the formation of an increasingly deep dip due to an attractive critical Casimir force, in agreement with the theoretical predictions. For clarity, symbols and curves corresponding to different temperatures have been separated vertically by a shift of $1\times k_{\rm B}T$.}
\label{fig:1}
\end{figure}

\renewcommand{\figurename}{{\bf Figure}}
\setcounter{figure}{1}
\renewcommand{\thefigure}{{\bf \arabic{figure}}}
\begin{figure}
\includegraphics[width=0.7\textwidth]{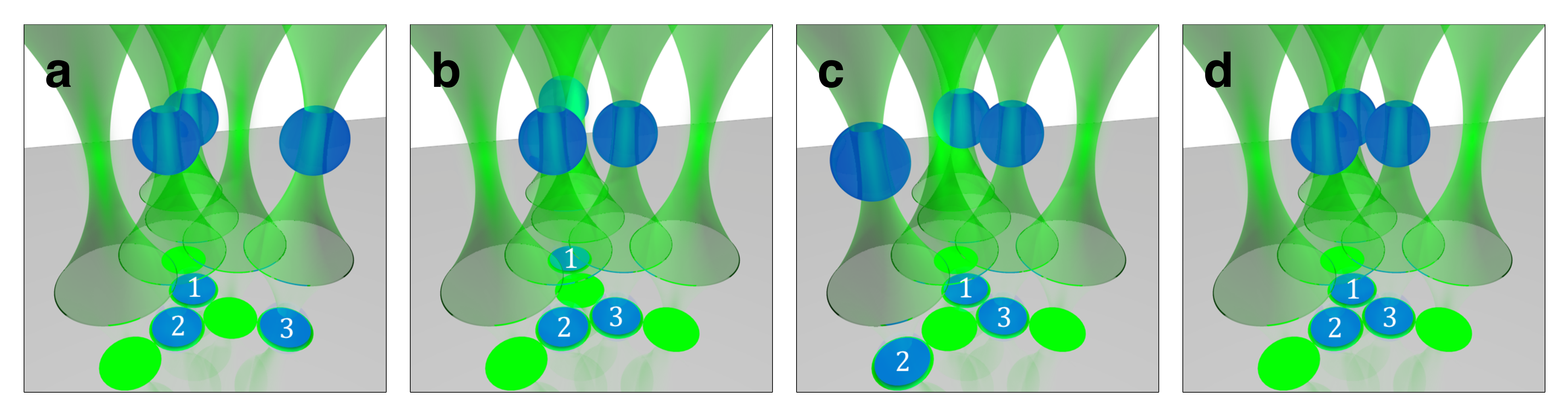}
\caption{
{\bf Experimental configurations for the measurement of many-body critical Casimir forces.}
Cartoon of the geometrical arrangement of the six optical tweezers (green conoids) and the colloids (blue spheres) during the experiment. The resulting harmonic optical traps are arranged as indicated beneath by the schematic orthogonal projection on the coverslip, with colloids and traps represented by numbered blue and dark green circles, respectively. The size of the particles are to scale with their distance, but not with the distance from the coverslit. At each temperature $T$,  the interactions between the pairs 1-2, 1-3 and 2-3 of colloids are measured in the configurations {\bf a}, {\bf b} and {\bf c}, respectively. The effective potential between colloids 1 and 2 in the presence of colloid 3 is then measured in the configuration {\bf d}. While measuring the pair interactions in {\bf a}-{\bf c}, the remaining colloid is optically moved into the nearest external trap. All six optical traps are always switched on during the measurements in order not to alter the hologram on the spatial light modulator and the corresponding optical potentials.}
\label{fig:2}
\end{figure}

\renewcommand{\figurename}{{\bf Figure}}
\setcounter{figure}{2}
\renewcommand{\thefigure}{{\bf \arabic{figure}}}
\begin{figure}
\includegraphics[width=.6\textwidth]{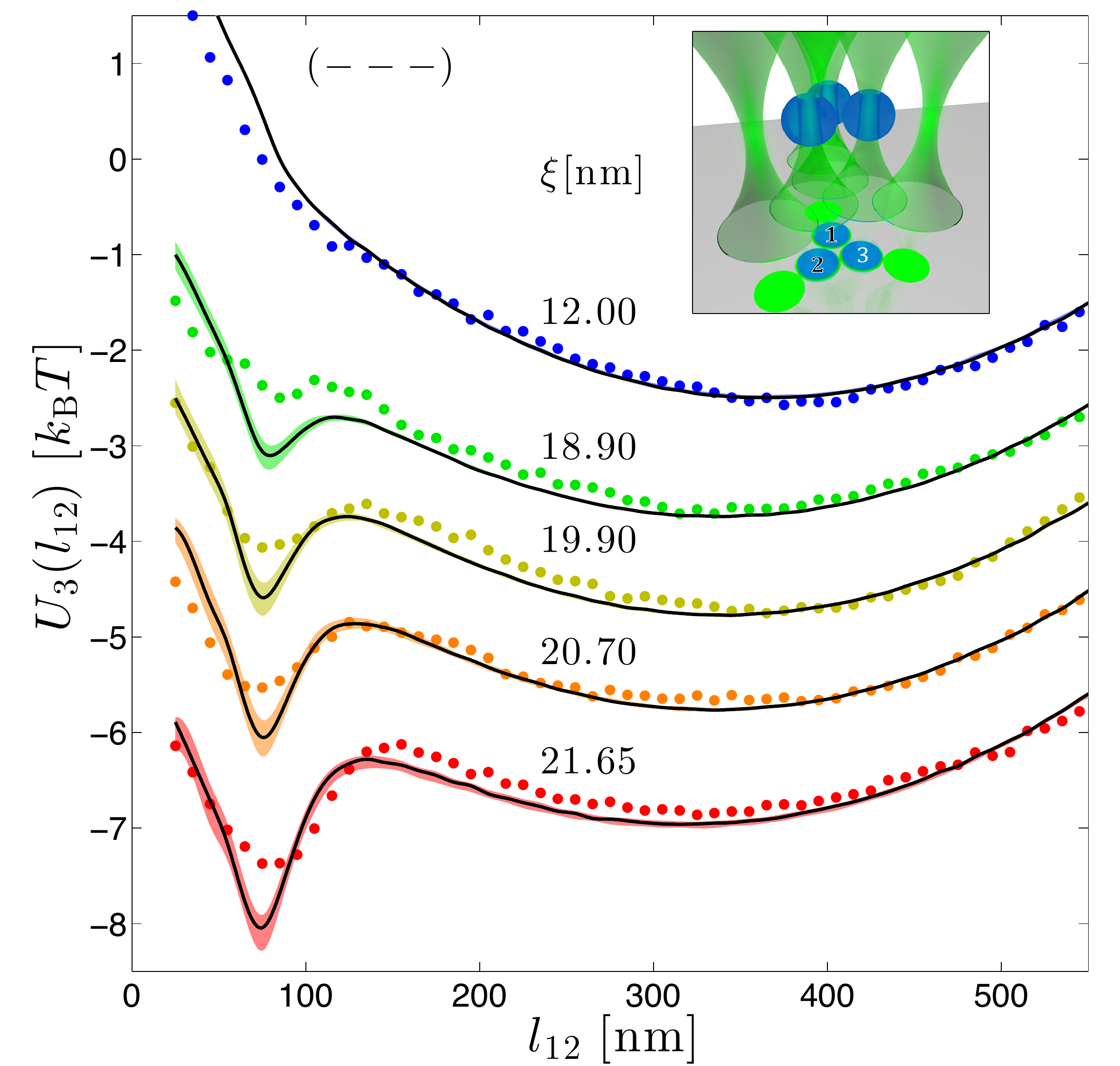}
\caption{
{\bf  Experimental evidence of many-body critical Casimir forces among hydrophilic particles.}
The symbols represent the measured effective potential $U_3(l_{12})$ between the particles 1 and 2 (labelled in black in the inset) in the presence of particle 3 (labelled in white in the inset) as a function of the in-plane surface-to-surface distance $l_{12}$ upon increasing (from top to bottom) the correlation length $\xi$. All particles are hydrophilic ($---$), resulting in attractive critical Casimir forces. The solid lines represent the corresponding theoretical predictions obtained by assuming additivity of the measured pair potentials between particles 1-2, 1-3 and 2-3 with the associated uncertainty indicated by the shading.  The observed discrepancy increases as $\xi$ increases, providing quantitative evidence of the nonadditive nature of the critical Casimir interactions. The colour code of the data points is the same as in Fig.~1, and symbols and lines are vertically separated by $1\times k_{\rm B}T$ for reasons of clarity.
Inset: Cartoon of the trap and colloid configuration during the measurement (see Fig.~2 for details).}
\label{fig:3}
\end{figure}

\renewcommand{\figurename}{{\bf Figure}}
\setcounter{figure}{3}
\renewcommand{\thefigure}{{\bf \arabic{figure}}}
\begin{figure}
\includegraphics[width=.6\textwidth]{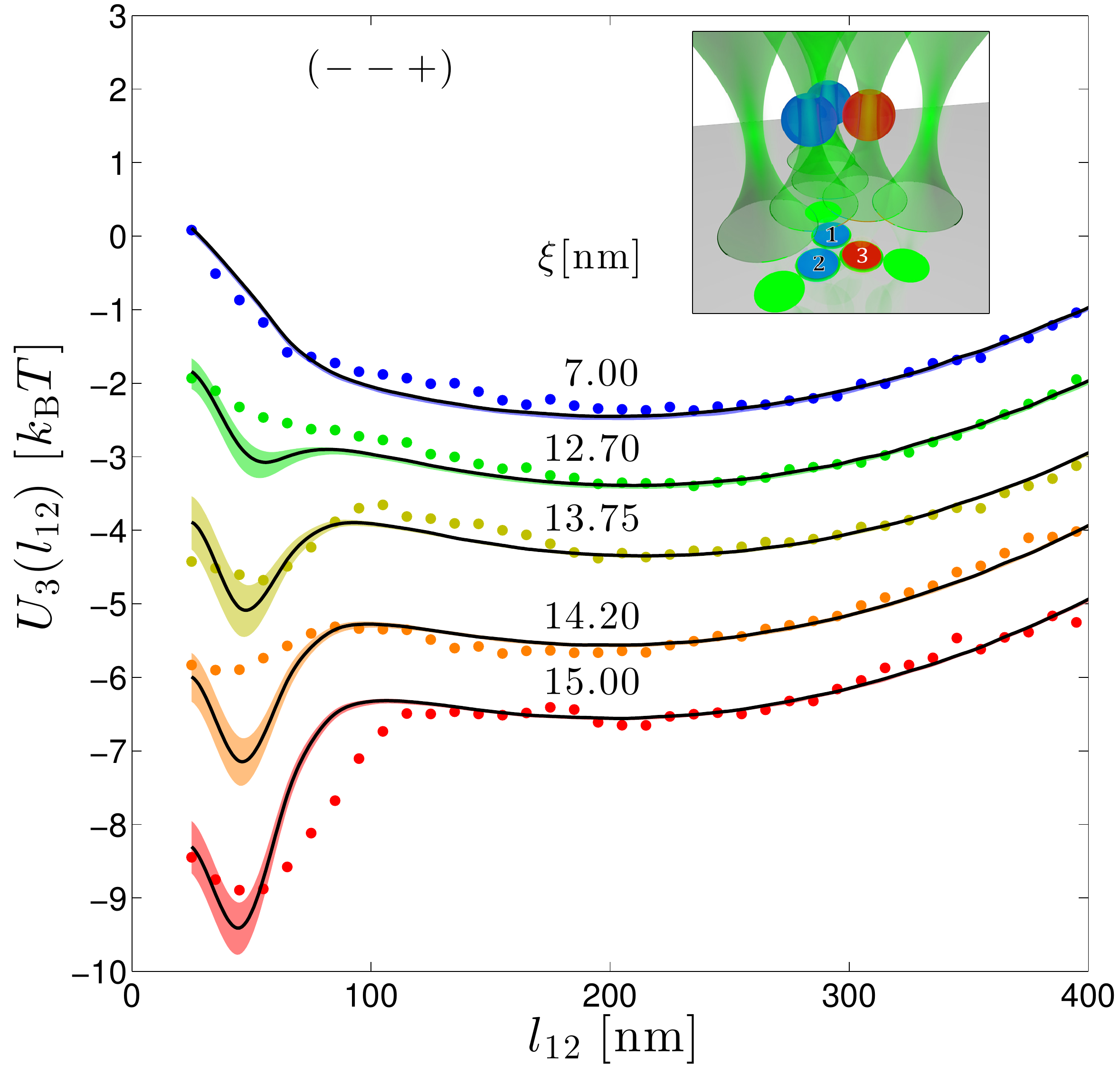}
\caption{
{\bf  Experimental evidence of many-body critical Casimir forces among one hydrophobic and two hydrophilic particles.}
The symbols represent the measured effective potentials $U_3(l_{12})$ between particles 1 and 2 (blue, labelled in black in the inset) in the presence of particle 3 (red, labelled in white in the inset) as a function of the in-plane surface-to-surface distance $l_{12}$ upon increasing (from top to bottom) the correlation length $\xi$. Particles 1 and 2 (blue spheres in the inset) are hydrophilic ($-$), while particle 3 (red sphere in the inset) is hydrophobic ($+$), so that the two-body critical Casimir forces cause attraction between 1 and 2 and repulsion between 2 and 3 and between 3 and 1. The solid lines represent the corresponding theoretical prediction obtained by assuming additivity of the measured pair potentials between particles 1-2, 1-3 and 2-3, with the associated uncertainty indicated by the shading. The observed discrepancy between the lines and the symbols increases as $\xi$ increases, providing quantitative evidence of the nonadditive nature of critical Casimir interactions also in the presence of opposing boundary conditions. Symbols and lines corresponding to different temperatures are vertically separated by $1\times k_{\rm B}T$ for reasons of clarity. Inset: Cartoon of the trap and colloid configuration during the measurement (see Fig.~2 for details).}
\label{fig:4}
\end{figure}

\renewcommand{\figurename}{{\bf Supplementary Figure}}
\setcounter{figure}{0}
\renewcommand{\thefigure}{{\bf \arabic{figure}}}
\begin{figure}
\includegraphics[width=.53\textwidth]{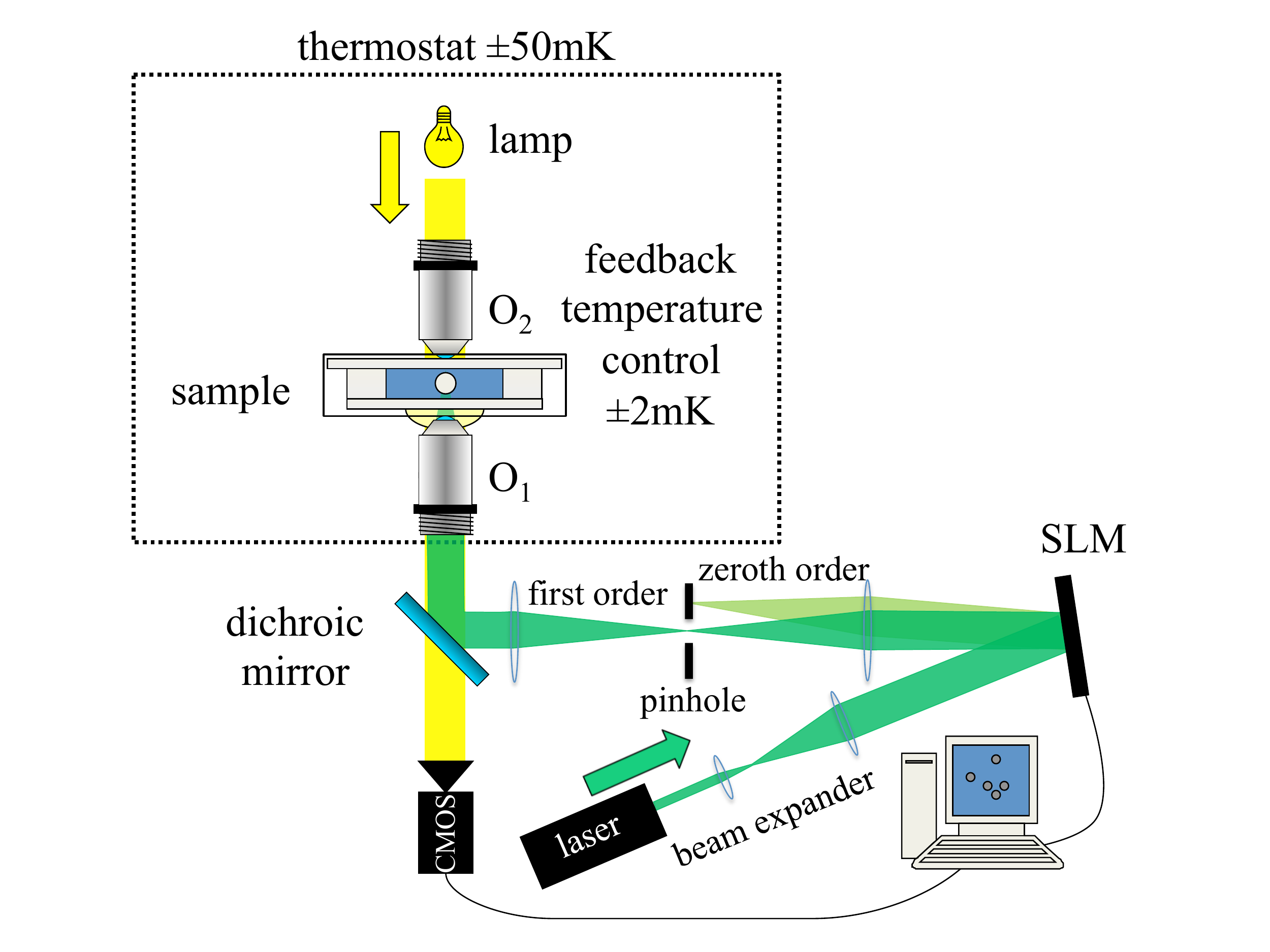}
\caption{
{\bf Experimental setup.}
The experimental setup consists of  holographic optical tweezers (HOTs) and a digital video microscope (DVM). The former generates multiple reconfigurable optical traps by imposing a phase-only hologram on an incoming beam using a spatial light modulator (${\rm SLM}$) and by focusing the resulting beam with a high-numerical-aperture objective (${\rm O_1}$). The latter, instead, tracks the position of the colloids within the horizontal $x$-$y$ plane with nanometric accuracy by illuminating the sample with white light through a condenser (${\rm O_2}$). A dichroic mirror (${\rm DM}$) is employed in order to combine the optical paths of the laser and of the white light. The temperature of the sample is controlled as explained in the Methods and as indicated in the schematic of the setup.}
\label{edf:1}
\end{figure}

\renewcommand{\figurename}{{\bf Supplementary Figure}}
\setcounter{figure}{1}
\renewcommand{\thefigure}{{\bf \arabic{figure}}}
\begin{figure}
\includegraphics[width=.38\textwidth]{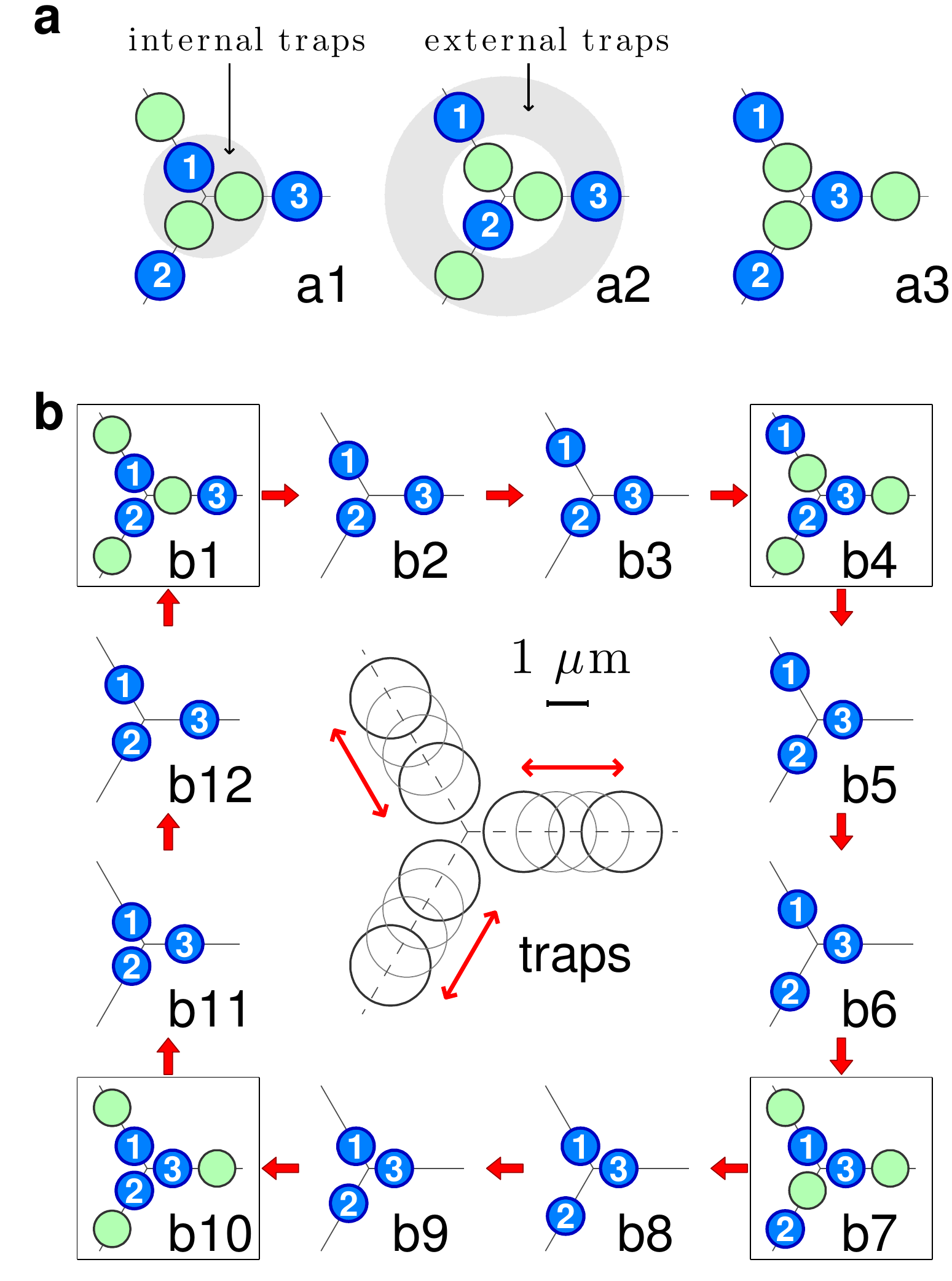}
\caption{
{\bf Particle configurations and protocol to switch between them.}
{\bf a}, Particle configurations for the characterisation of the one-body optical potential
associated with each optical trap. Empty and occupied optical traps are represented by green and blue circles, respectively.
{\bf b}, Protocol to switch between the configurations (framed) in which the measurements are eventually performed: at each temperature, first the two-body interactions between the pairs 1-2, 1-3 and 2-3 of colloids are measured in the configurations b1 (Fig.~2a), b4 (Fig.~2b) and b7 (Fig.~2c), respectively; then, the effective potentials between the three colloids is measured in the configuration b10 (Fig.~2d). For rearranging the colloids in these configurations, the intermediate optical traps indicated by thin grey circles in the central sketch (with the associated scale) are switched on and off in order to move the particles from the internal to the external traps and vice versa, realising the intermediate configurations indicated in the panel. }
\label{edf:2}
\end{figure}

\renewcommand{\figurename}{{\bf Supplementary Figure}}
\setcounter{figure}{2}
\renewcommand{\thefigure}{{\bf \arabic{figure}}}
\begin{figure}
\includegraphics[width=.75\textwidth]{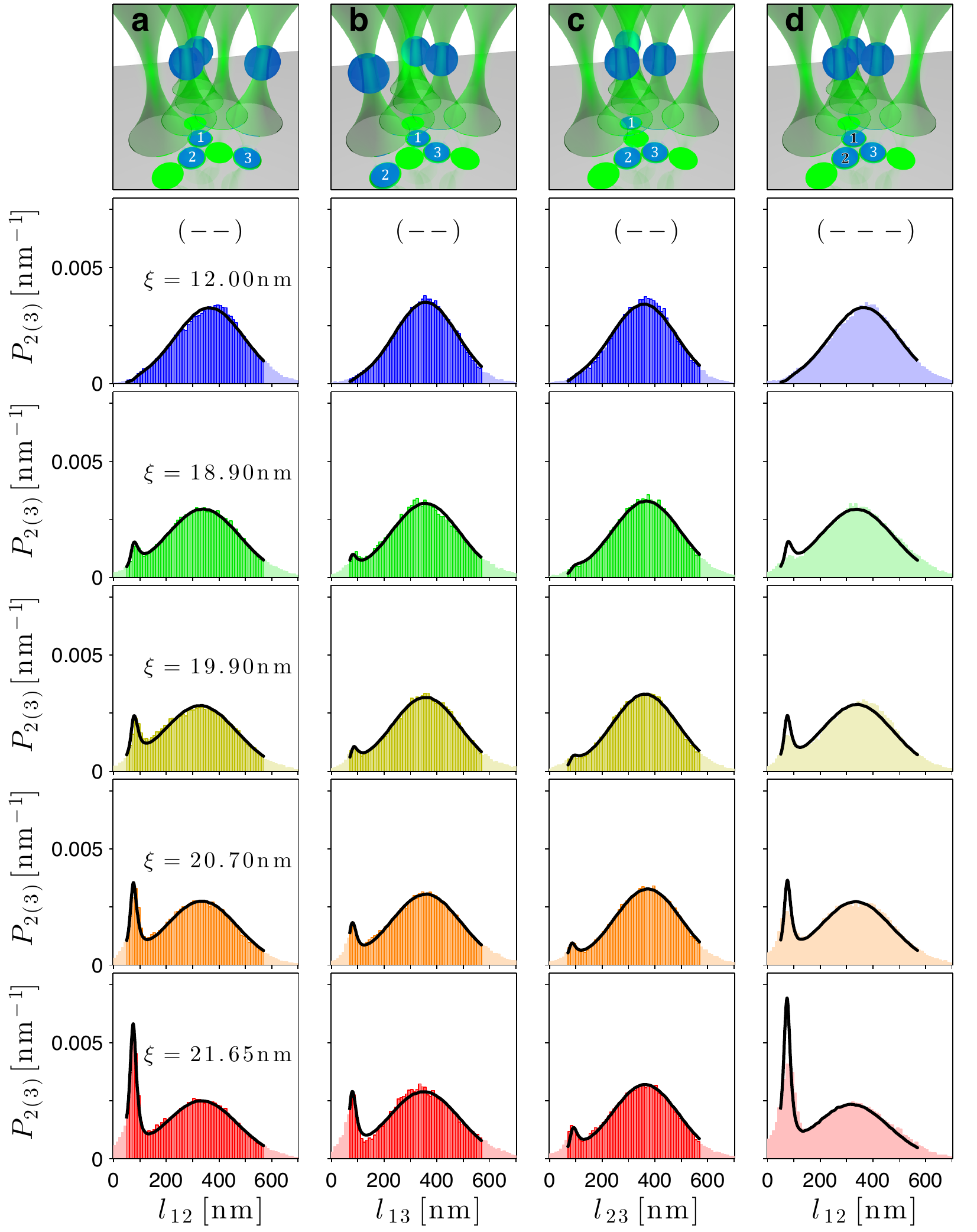}
\caption{
{\bf Interparticle separation histograms for three hydrophilic particles ($---$).}
{\bf a}-{\bf c}, Histograms of the in-plane surface-to-surface distance $l_{ij}$
between particles 1-2, 1-3 and 2-3, shown in columns {\bf a}, {\bf b} and {\bf c}, respectively, for increasing values of the correlation length $\xi$, from top to bottom; the values given for $\xi$ hold for the whole corresponding line of panels. The solid lines represent fits  to the theory corresponding to Eq.~(\ref{eq:U2}), which are very good. For each row of plots, $\xi$ was determined by the best global fit to the experimental data for $P_2(l_{ij})$ highlighted by darker colours in the two-particle configurations {\bf a}, {\bf b} and {\bf c}. The remaining parameters were already determined at low temperatures, at which the critical Casimir potential is negligible.
{\bf d}, Histograms of the in-plane surface-to-surface distance $l_{12}$ between particle 1 and 2 (labelled in black)  in the presence of particle 3 (labelled in white). The solid lines indicate the theoretical prediction assuming additive critical Casimir forces and show a clear discrepancy (mainly at the left peak) with the experimental histograms, which increases upon increasing $\xi$. The histograms in columns {\bf a} and {\bf d} correspond to the effective potentials reported in Fig.~1c and 3, respectively.
First row: Cartoon of the trap and colloid configurations during the measurement (see Fig.~2 for details).}
\label{edf:3}
\end{figure}

\renewcommand{\figurename}{{\bf Supplementary Figure}}
\setcounter{figure}{3}
\renewcommand{\thefigure}{{\bf \arabic{figure}}}
\begin{figure}
\includegraphics[width=.75\textwidth]{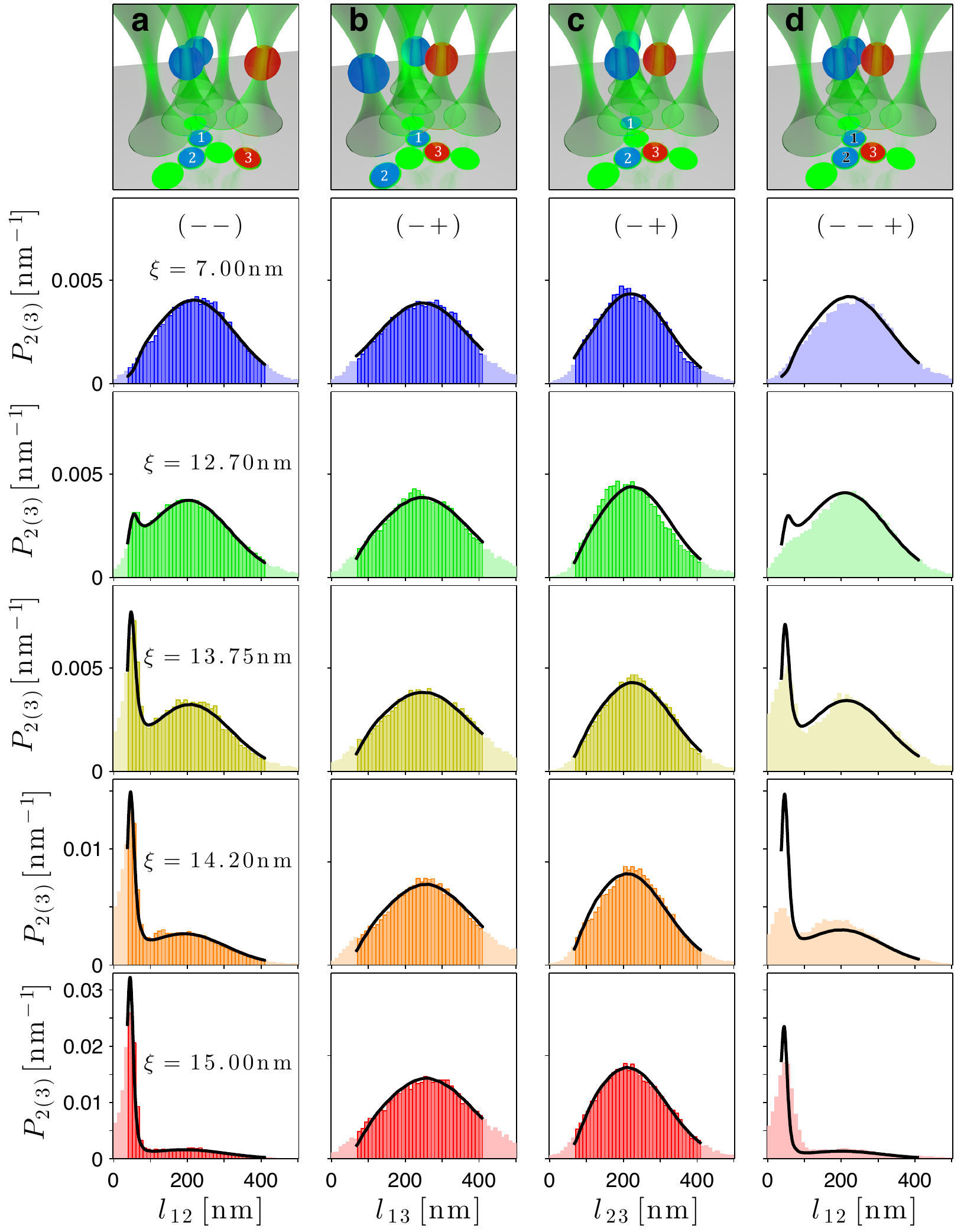}
\caption{
{\bf Interparticle separation histograms for two hydrophilic particles and one hydrophobic particle ($--+$).}
{\bf a}-{\bf c},  Histograms of the in-plane surface-to-surface distance $l_{ij}$ between particles 1-2 ($--$), 1-3 ($-+$) and 2-3 ($-+$), shown in columns {\bf a}, {\bf b} and {\bf c}, respectively, for increasing values of the correlation length $\xi$, from top to bottom; the values given for $\xi$ hold for the whole corresponding line of panels. The solid lines represent fits to the theory corresponding to Eq.~(\ref{eq:U2}), which are very good. As in Supplementary Fig.~3,  $\xi$ was determined for each row of plots by the best global fit to the experimental data for $P_2(l_{ij})$ highlighted by darker colours in the two-particle configurations {\bf a}, {\bf b} and {\bf c}. The remaining parameters were already determined at low temperatures, at which the critical Casimir potential is negligible. Note that the histograms in columns {\bf b} and {\bf c} do not feature the emergence of a pronounced peak (corresponding to an attractive dip in the effective potential) because the critical Casimir pair potential with ($-+$) boundary conditions is repulsive.
{\bf d}, Histograms of the in-plane surface-to-surface distance $l_{12}$ between particles 1 and 2 ($--$, blue spheres) in the presence of particle 3 ($+$, red sphere). The solid lines show the theoretical predictions assuming additive critical Casimir forces; there are clear discrepancies with the experimental histograms, which become more pronounced upon increasing $\xi$. The histograms in {\bf d} correspond to the effective potentials reported in Fig.~4. First row: Cartoon of the trap and colloid configurations during the measurement (see Fig.~2 for details), with hydrophilic and hydrophobic particles represented as blue and red spheres, respectively.}
\label{edf:4}
\end{figure}

\end{widetext}

\end{document}